\documentclass[journal]{IEEEtran}
\usepackage{cite}
\usepackage{amsmath,amssymb,amsfonts}
\usepackage{algorithmic}
\usepackage{graphicx}
\usepackage{textcomp}
\usepackage[label font=bf,labelformat=simple]{subfig}
\usepackage{floatrow}
\usepackage{wrapfig}
\newtheorem{Remark}{Remark}
\newtheorem{Proposition}{Proposition}
\newtheorem{Problem}{Problem}
\newtheorem{myTheo}{Theorem}
\def\BibTeX{{\rm B\kern-.05em{\sc i\kern-.025em b}\kern-.08em
    T\kern-.1667em\lower.7ex\hbox{E}\kern-.125emX}}
\begin{document}
\title{A Frequency-Domain Path-Following Method for Discrete Data-Based Paths}
\author{Zirui Chen and Zongyu Zuo,~\IEEEmembership{Senior Member,~IEEE}
\thanks{This work was supported by the National Natural Science Foundation of China under Grant 62073019.}
\thanks{Z. Chen and Z. Zuo are with Seventh Research Division, Beihang University (BUAA), Beijing,
 100191, China.  E-mail: chenzirui@buaa.edu.cn; zzybobby@buaa.edu.cn.}}

 \markboth{IEEE TRANSACTIONS ON AUTOMATIC CONTROL}%
 {Shell \MakeLowercase{\textit{et al.}}: Bare Demo of IEEEtran.cls for IEEE Journals}
 
\maketitle

\begin{abstract}
    This paper presents a novel frequency-domain approach for path following problem, specifically designed to handle paths described by discrete data. The proposed algorithm utilizes the fast Fourier Transform (FFT) to process the discrete path data, enabling the construction of a non-singular guiding vector field. This vector field serves as a reference direction for the controlled robot, offering the ability to adapt to different levels of precision. Additionally, the frequency-domain nature of the vector field allows for the reduction of computational complexity and effective noise suppression. The efficacy of the proposed approach is demonstrated through a numerical simulation, and theoretical analysis provides an upper bound for the ultimate mean-square path-following error.

\end{abstract}


\section{Introduction}
The path following problem is a fundamental issue in the domain of robot control. Distinguishing itself from the trajectory tracking problem, the primary focus of the path following problem lies in pinning the position of the robot onto a desired geometric curve rather than a particular point at a given moment in time. This application has been extensively employed in various path following scenarios such as $3D$ path following \cite{Yao_2020Path}, moving path following \cite{Wang_2019Cooperative,Oliveira_2016}, and coordinated path following \cite{Zuo_2022Coordinated}, among others. A guiding vector field approach has proven to be effective in tackling these problems from a global perspective \cite{Sujit_2014Unmanned}. Notably, a general approach for constructing guiding vectors for time-varying curves in $n$-dimensional space has been proposed \cite{Goncalves_2010}, leading to significant developments in singularity-free guiding vector fields \cite{Yao_2021Singularity}, constructive time-varying vector fields \cite{Rezende_2022Constructive}, and guiding vector fields on Riemannian manifolds \cite{Yao_2023Topological}.

Although the guiding vector field method is widely applicable for solving path following problems, the desired curves typically discussed in literature are characterized by a specific mathematical expression, such as $\alpha(x,y) = 0$. However, in practical scenarios, it can be challenging to provide a precise mathematical expression for the desired path. Moreover, measurement data can be affected by noise, further complicating the problem. To the best of our knowledge, there have been limited publications addressing the specific topic of data-based path following. However, in the realm of data-based control problems, there has been a notable increase in the number of recent works that have emerged, such as the stochastic distribution control problem \cite{9373984,9780596,9362158}, the data-driven control problem \cite{9484756,10101826,10122597,10107903}, the model-free control problem \cite{9345499}, and so on. Different from model-based control, data-driven control is a control strategy that relies on empirical data rather than explicit mathematical models to derive control laws. This approach can be advantageous when the system being controlled is complex, nonlinear, or poorly understood, and a mathematical model of the system may be difficult or impossible to formulate. One key advantage of data-driven control is its adaptability to changes or uncertainties in the system, as it can adjust its control strategies based on new data.

The discussion above has shown that data-driven approaches can facilitate the use of path-following control scheme in practical applications, where the desired path may be descried by discrete data or may be subject to measurement errors. Therefore, this paper proposes a non-singular guided vector field using Fast Fourier Transform based on path data. This vector field can navigate the target from a global data perspective, avoiding singularities in the process. Specifically, the main contributions can be outlined as follows:
  \begin{enumerate}
    \item A globally defined non-singular path following algorithm is proposed for paths represented by discrete data. This algorithm is based on a non-singular guiding vector field using the Fast Fourier Transform results from the original path data, thus formulating a frequency domain approach for path-following problem.
    \item This algorithm exhibits certain capabilities in suppressing measurement errors. Accordingly, this paper provides an estimate of the ultimate mean-square upper tracking error bound, as defined in Section~\ref{s32}, to evaluate the performance of the path-following process.
    \item The data processing utilizes Fast Fourier Transform, and the algorithm can control the width of the window function $m$ to adjust the computational complexity and precision based on the performance of the controller.
 \end{enumerate}

 The subsequent sections of this paper are structured as follows. In Section \ref{section2}, the fundamentals of epicycles and guiding vector fields for planar path following are introduced. The first part of Section \ref{section3} elucidates the extension of the guiding vector field for data-based desired paths, while the second part delves into the impact of a rectangle window and the Gauss noise in this regard. Additionally, Section \ref{section4} details a simulation wherein a $\pi$ pattern composed of $758$ data points serves as the desired path. Lastly, concluding remarks are expounded upon in Section \ref{section5}.

\section{Preliminaries}
\label{section2}
\subsection{Notation}
 The frame in this brief paper is $\mathbb{R} ^2$. The notation `:=' refers to `is defined as'. The set $Z_N$ represents a set $\{n\leq N | n \in N^*\}$ where $N^*$ is the set of non-negative integers and the set $\bar{Z}_N = \{-\frac{N}{2}\leq n\leq \frac{N}{2} | n \in Z\}$ when $N$ is even and $\bar{Z}_N = \{-\frac{N-1}{2}\leq n\leq \frac{N-1}{2} | n \in Z\}$ when $N$ is odd. The set minus is defined as $A \setminus B := \{\alpha|\alpha \in A \ \text{and} \  \alpha \notin B\}$. For a smooth function $f: \mathbb{R}^n \rightarrow \mathbb{R} $, its gradient is defined as $\nabla f = \begin{bmatrix} \frac{f}{\partial x^1} , \ldots, \frac{f}{\partial x^n} \end{bmatrix}^{\top}$ where $x^1,\ldots,x^n$ serve as the coordination of $\mathbb{R}^n$.

\subsection{Epicycle}
Epicycle is an ancient concept in astronomy used in explaining the observed motion of planets in the geocentric model of the universe. The astronomers in ancient time used to describe the orbit of a single planet as a series of epicycles superimposed upon one another. 

Despite having been abandoned for centuries, the idea of epicycles regained relevance in modern times following Joseph Fourier's (1768-1830) introduction of the Fourier series method, which demonstrated the ability to decompose a periodic orbit into a Fourier series and gave new significance to the concept of epicycles.  

\begin{figure}
    \centering
    \includegraphics{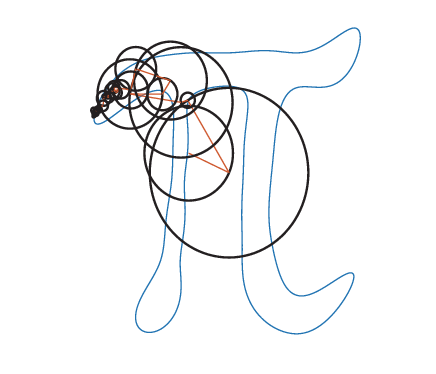}
    \caption{The orbit of $\pi$ approached by $40$ epicycles}
    \label{pi_epicycle}
\end{figure}

\begin{figure} [!ht]
	\centering
	\subfloat[\label{fig:2a}]{
		\includegraphics[width=0.28 \linewidth]{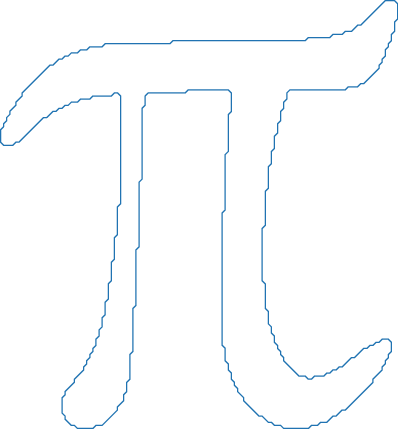}}
	\subfloat[\label{fig:2b}]{
		\includegraphics[width=0.33 \linewidth]{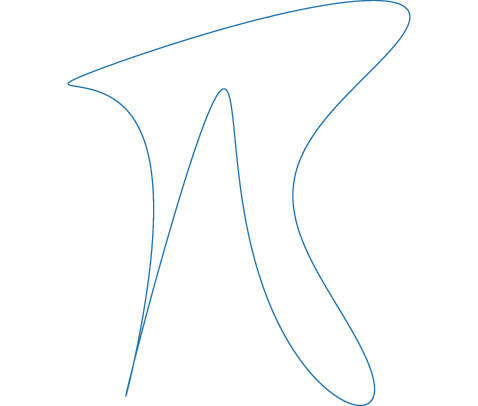}}
	\subfloat[\label{fig:2c}]{
		\includegraphics[width=0.30 \linewidth]{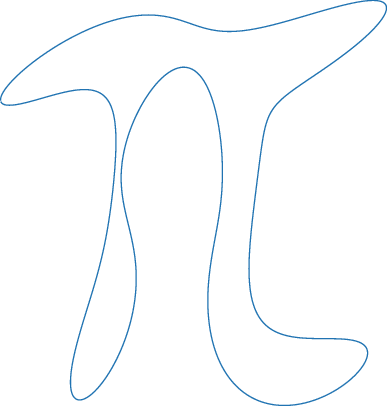} } \\
	\subfloat[\label{fig:2d}]{
		\includegraphics[width=0.33 \linewidth]{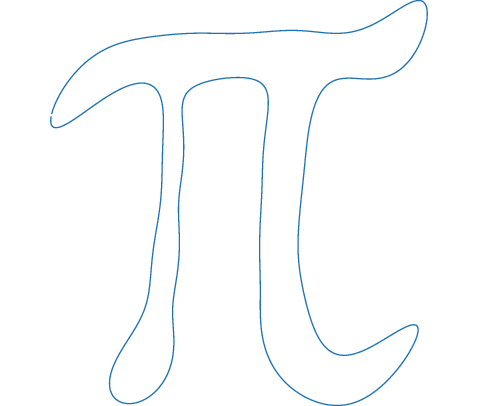}}
    \subfloat[\label{fig:2e}]{
        \includegraphics[width=0.33 \linewidth]{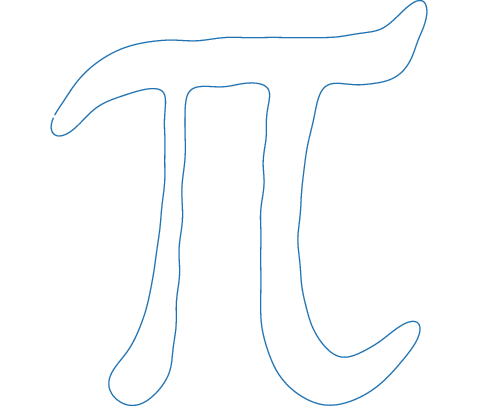}}
    \subfloat[\label{fig:2f}]{
        \includegraphics[width=0.28 \linewidth]{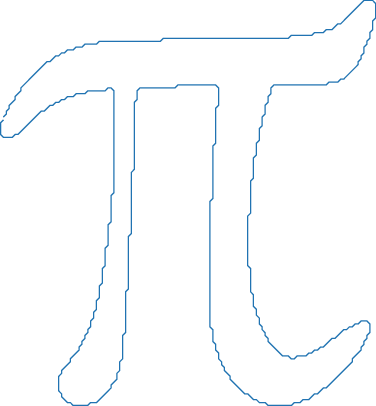}}
	\caption{A $\pi$ pattern described by $758$ sample points and the reconstructed patterns via different number of epicycles. (\ref{fig:2a}): the original pattern. (\ref{fig:2b})-(\ref{fig:2f}) are the reconstructed patterns with 10, 20, 40, 100, 758 epicycles, respectively}
	\label{diff pi epi}
\end{figure}

With discrete Fourier Transformation, every orbit described by a series of data could be approached with certain epicycles (Fig.\ref{pi_epicycle} for example).\\
 Generally speaking, the low-frequency component of the Fourier series approximates the overall shape of the original curve, while the high-frequency component provides a more intricate representation of its specific details (see Fig.\ref{diff pi epi}). In practical applications, eliminating some high-frequency terms does not hinder the ability of the remaining low-frequency terms to efficiently reconstruct the curve. Furthermore, the introduction of this cutoff (commonly referred to as the window function in signal processing) effectively diminishes high-frequency noise commonly found in signal measurements. This, coupled with the implementation of fast Fourier transforms that offer a lower complexity approach to calculating Fourier series, undeniably establishes the method's significant value in engineering applications.

\subsection{General Guiding Vector Field for 2-D path}
\label{section24}
For a 2-D path $\alpha(x,y) = 0$, the purpose of a guiding vector field $\chi$ is to arrange every flow $\xi(t)\in \mathbb{R}^2$ with the dynamics governed by $\dot{\xi}(t) = \chi(\xi(t))$ to ultimately converges towards the curve $\alpha = 0$ while $\dot{\xi}(t) \neq 0$. \\
It has shown in \cite{Goncalves_2010} that a guiding vector field with the following form (\ref{singular vector field}) could reach the goal:
\begin{equation}
    \label{singular vector field}
    \chi = -G \nabla V + H (\wedge \nabla \alpha)
\end{equation}
where a candidate Lyapunov function $V = \frac{1}{2}\alpha^2$ is chosen. $G$ and $H$ are constant diagonal positive definite matrix and $\wedge \nabla \alpha$ is a certain vector showing the property $\nabla \alpha \cdot \wedge \nabla \alpha =0$. One obtains that for system $\dot{\xi}(t) = \chi(\xi(t))$, the following equation holds:
\begin{equation}
    \dot{V} = -G\alpha^2 \Vert \nabla \alpha \Vert^2
\end{equation}
Thus, $\alpha$ would converge to zero. And when $\alpha = 0$, $\chi = H (\wedge \nabla \alpha)$, indicating that $\chi$ is not always zero vector on the desired curve. \\
\indent Although the analysis above has concluded that (\ref{singular vector field}) could reach the goal of converging and moving around the curve, two limitations ought to state here:
\begin{enumerate}
    \item The expression (\ref{singular vector field}) can not guarantee itself always non-zero, deriving singular points caused by zero vector.
    \item The vector field (\ref{singular vector field}) works only when the curve has the form $\alpha(x,y)=0$. While for certain curve ($\pi$ in Fig. \ref{pi_epicycle} for example), it is nearly impossible to find an expression like $\alpha(x,y) = 0$.
\end{enumerate}
\subsection{Problem Formulation}\label{subsec:PF}
In situations where it is difficult to express the path using a simple expression such as $\alpha(x,y)=0$, an alternative approach is to represent the path as a finite series of sample data points, denoted by $x[n]$ and $y[n]$. Besides, it is worth noting that in practical engineering applications, this data may be subject to noise, which complicates the accurate determination of the path. Consequently, the primary issue addressed in this article may be formulated as follows.:
\begin{Problem}
    Given desired path described by data $\{x_d[n],y_d[n]\}$, design a guiding vector field $\chi$ based on the data $\{\tilde{x}_d[n],\tilde{y}_d[n]\}$, which is composed of data $\{x_d[n],y_d[n]\}$ and Gauss noise $\mathcal{N}(0,\sigma^2)$, satisfying the following requirements:
    \begin{enumerate}
        \item No zero vector would be contained in this vector field $\chi$.
        \item For dynamical system $\dot{\xi}(t) = \chi(\xi(t))$, $\xi(t) \in \mathbb{R}^2$, its trajectory would converge to the desired path (reconstructed by the given data $\{x_d[n],y_d[n]\}$) with an error upper bounded by $\delta$ of mean-square meaning.
    \end{enumerate}
\end{Problem}
\section{Main Results}
\label{section3}
Within this section, our focus will be on addressing the primary problem outlined in \emph{Section \ref{subsec:PF}}. Initially, our approach will aim to derive a vector field that is non-singular and guides to the precise path in cases where data is free from any noise interference. Subsequently, we will incorporate the impact of noise, and introduce a rectangular window function as a filtering mechanism to reject the noise and expedite the vector field generation process.
\subsection{Generation of Guiding Vector Field}
\label{section31}
In this subsection, the path we discussed is given by $N$ sampling points on it. The purpose is to derive a non-singular vector field
\begin{equation}
    p[n] = \left(x_d[n] , y_d[n]\right)^{\top} \  n\in Z_N
\end{equation}
\indent Given that these points lie on a plane, it is possible to characterize them by means of complex numbers.
\begin{equation}
    c[n] = x_d[n] + \mathrm{j} y_d[n] \ \  n\in Z_N
\end{equation}
where $\mathrm{j}^2 = -1$. Thus, the following discrete Fourier Transformation holds
\begin{equation}
    \label{DFT}
    \left\{\begin{aligned}
        c[n] &= \sum_{k \in Z_N} a_k e^{\mathrm{j}k\frac{2\pi}{N}n} \\
        a_k & = \frac{1}{N} \sum_{k\in Z_N} c[n] e^{-\mathrm{j}k\frac{2\pi}{N}n}
    \end{aligned}\right.
\end{equation}
\indent Rewrite the first equation in (\ref{DFT}) in an $\mathbb{R}^2$ form
\begin{equation}
    \label{planar points}
    \left\{\begin{aligned}
            x_d[i] &=  \sum_{k\in Z_N} \Vert a_k \Vert \cos{\Big(\frac{2\pi i k}{N}+\mathrm{arg}(a_k)\Big)} \\
            y_d[i] &=  \sum_{k\in Z_N} \Vert a_k \Vert \sin{\Big(\frac{2\pi i k}{N}+\mathrm{arg}(a_k)\Big)}
        \end{aligned} \right.
\end{equation}
where $i\in Z_N$ and $\mathrm{arg}(\cdot)$ is the principal value of argument referring to the unique argument value of a complex number. To get a continuous curve, a new parameter $\theta := \frac{2\pi i}{N}$. From this perspective, $x[i]$ and $y[i]$ could be regarded as sampling points of smooth curve $x(\theta)$ and $y(\theta)$ with expressions
\begin{equation}
    \label{x=andy=}
    \left\{\begin{aligned}
        x_d(\theta) &=  \sum_{k\in \bar{Z}_N} \Vert a_k \Vert \cos{(k\theta+\mathrm{arg}(a_k))} \\
        y_d(\theta) &=  \sum_{k\in \bar{Z}_N} \Vert a_k \Vert \sin{(k\theta+\mathrm{arg}(a_k))}
    \end{aligned}\right.
\end{equation}
where $a_k$ could be derived from a fast Fourier transformation.
\begin{Remark}
    It is noteworthy to emphasize the fact that the index sets for $k$ in (\ref{planar points}) and (\ref{x=andy=}) are different. This distinction is caused by a critical frequency domain property of Discrete Fourier Transform (DFT). In this paper, this property is $a_{N-k} = a_{-k}$, which is applicable only in the discrete scenario. To obtain the interpolation using DFT outcomes, the Fourier coefficients must be shifted into the negative domain.
\end{Remark}
\begin{Proposition}
    \label{Proposition1}
    $c(\theta) = x_d(\theta)+\mathrm{j}y_d(\theta)$ with $x_d(\theta)$ and $y_d(\theta)$ obtained from (\ref{x=andy=}) is an interpolation for (\ref{planar points}) holding the following properties:
    \begin{itemize}
        \item $c(\frac{2\pi n}{N}) = c[n]$.
        \item $c(\theta)$ is a smooth complex function.
    \end{itemize}
\end{Proposition}
\indent In order to build a non-singular guiding vector field, the original path (\ref{x=andy=}) ought to be rewritten in the form of
\begin{equation}
    \label{parametric curve}
    \left\{\begin{aligned}
        \phi_1 &= x(t) -  \sum_{k\in \bar{Z}_N} \Vert a_k \Vert \cos{(k\theta+\mathrm{arg}(a_k))}\\
        \phi_2 &= y(t) -  \sum_{k\in \bar{Z}_N} \Vert a_k \Vert \sin{(k\theta+\mathrm{arg}(a_k))}
    \end{aligned}\right.
\end{equation}

\begin{myTheo}
    \label{theorem1}
    A non-singular guiding vector field to $\phi_1=0$ and $\phi_2=0$ is
    \begin{equation}
        \label{vector field origin}
        \chi = \nabla \phi_1 \times \nabla \phi_2 - k_1\phi_1 \nabla \phi_1 - k_2\phi_2 \nabla \phi_2
    \end{equation}
    where $k_1$ and $k_2$ are positive constants. (\ref{vector field origin}) satisfies the following two properties:
    \begin{enumerate}
        \item For $\eta(t) : \mathbb{R}^* \rightarrow \mathbb{R}^3$, where $\eta :=\begin{bmatrix} x(t) & y(t) &\theta(t) \end{bmatrix}^{\top} $, undergoes a dynamics of $\dot{\eta}(t) = \chi(\eta(t))$, the trajectory would finally converge to $\{\phi_1=0 \ \mathrm{and}\ \phi_2=0\}$,
        \item For every $\eta\in \mathbb{R}^3$, $\chi(\eta)\neq 0$.
    \end{enumerate}
\end{myTheo}
\begin{IEEEproof}
    From (\ref{vector field origin}), one obtains
    \begin{equation}
        \label{vector field expression}
        \chi = \begin{bmatrix}
            -\frac{\partial \phi_1}{\theta} \\ -\frac{\partial \phi_2}{\theta} \\ 1
        \end{bmatrix}
        -k_1 \phi_1 \begin{bmatrix}
            1 \\ 0 \\ \frac{\partial \phi_1}{\theta}
        \end{bmatrix}
        -k_2 \phi_2 \begin{bmatrix}
            0 \\ 1 \\ \frac{\partial \phi_2}{\theta}
        \end{bmatrix}
    \end{equation}
    \indent If (\ref{vector field expression}) becomes zero vector, it should hold $\phi_1 = -\frac{\partial \phi_1}{\partial \theta}$ and $\phi_2 = -\frac{\partial \phi_2}{\partial \theta}$, meaning the first two rows of (\ref{vector field expression}) are zeros. Noticing the fact that both $\phi_1$ and $\phi_2$ are real functions, thus the last row of (\ref{vector field expression}) becomes
    \begin{equation}
        1-\phi_1 \frac{\partial \phi_1}{\partial \theta} - \phi_2 \frac{\partial \phi_2}{\partial \theta} = 1+\phi_1^2+\phi_2^2 >0
    \end{equation}
    \indent Indicating that $\chi \neq 0$. Therefore, the second property holds. \\
    \indent To prove the first property, a candidate Lyapunov function is chosen as
    \begin{equation*}
      V_1 = k_1 \phi_1^2+k_2 \phi_2^2
    \end{equation*}
    where $k_1$ and $k_2$ are two positive constants. The time derivative $V_1$ is
    \begin{equation}
        \dot{V}_1 = -k_1^2 \phi_1^2 \nabla \phi_1^2 -k_2^2 \phi_2^2 \nabla \phi_2^2 \leq 0
    \end{equation}
    \indent According to \emph{Theorem 4.10} in \cite{khalil2002nonlinear}, $\dot{\xi}(t) = \chi(\xi(t))$ would converge to its invariant set, which is $\{\phi_1=0 \ \mathrm{and}\ \phi_2=0\}$. Property 1 holds.
\end{IEEEproof}
\begin{Remark}
    The non-singular guiding vector field (\ref{vector field origin}) is first proposed in \cite{Yao_2021Singularity}. The main idea behind this vector field is to transform the original closed paths (homeomorphic to a one-dimensional manifold $S$) into open paths (homeomorphic to a one-dimensional manifold $\mathbb{R}$) by dimensional expansion. This vector field does not possess singular points when applied to the latter case of the target path \cite{Yao_2021Singularity}. The paths reconstructed through Fourier transform, which are the focus of this study, precisely fall into the category of non-singular vector field research for the reason that the extra parameter $\theta$ as defined in this section could be viewed as the third dimension transforming the original path into a path homeomorphic to $\mathbb{R}$. Therefore, this paper employs such a guiding vector field as a method for path-following.
\end{Remark}
\subsection{Vector Field with a rectangular window}
\label{s32}
\begin{figure*} [t!]
	\centering
	\subfloat[\label{fig:3a}]{
		\includegraphics[width=0.33 \linewidth]{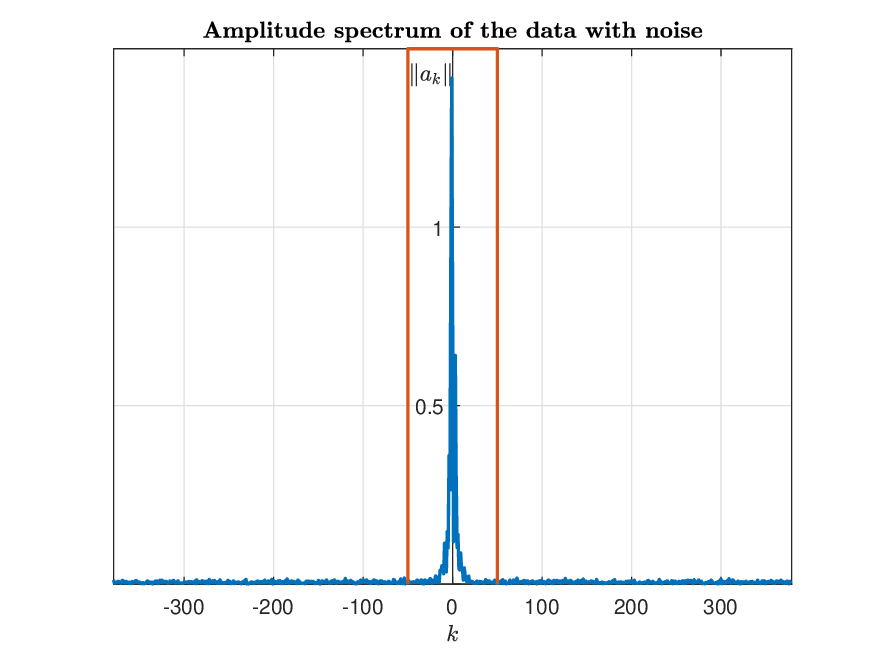}}
	\subfloat[\label{fig:3b}]{
		\includegraphics[width=0.33 \linewidth]{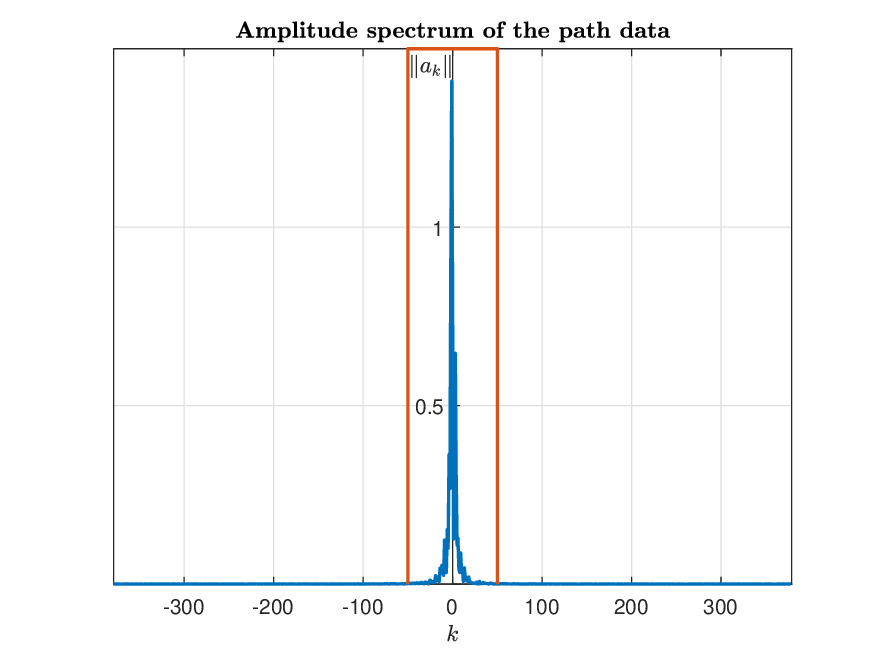}}
	\subfloat[\label{fig:3c}]{
		\includegraphics[width=0.33 \linewidth]{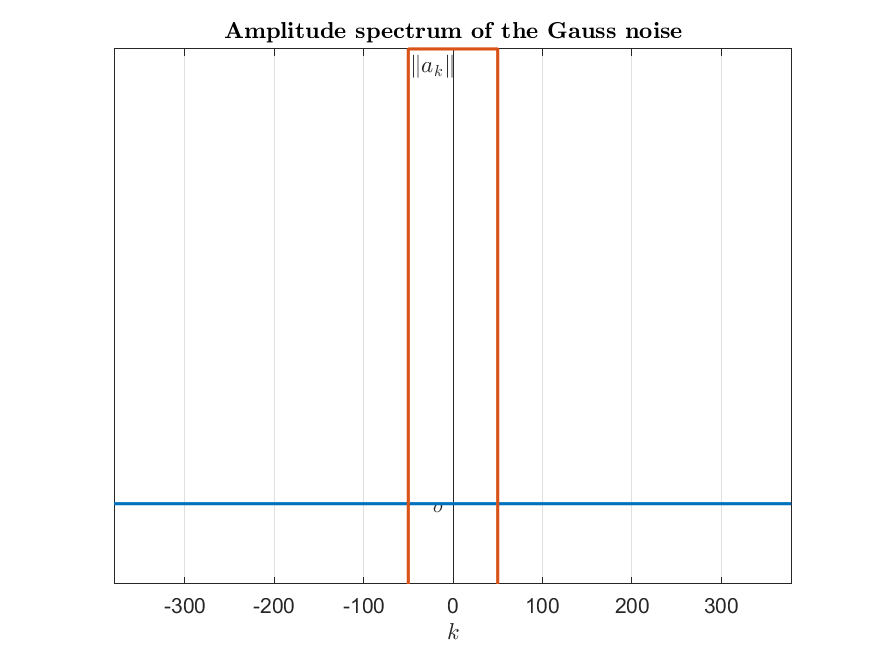} } \\
	\caption{ (\ref{fig:3a}): Amplitude spectrum of the data with noise. (\ref{fig:3b}): Amplitude spectrum of the path data. (\ref{fig:3c}): Amplitude spectrum of the Gauss noise }
	\label{fig3}
\end{figure*}

Based on the guiding vector field (\ref{vector field expression}), this subsection takes the noise into consideration and introduces a rectangular window working as a filter to restrict the impact of the noise. \\
\indent Owning to Proposition \ref{Proposition1}, (\ref{x=andy=}) is regarded as the precise expression of the desired curve. Namely, define $x(\theta)$ and $y(\theta)$ in (\ref{x=andy=}) as $x_d(\theta)$ and $y_d(\theta)$. This subsection takes into account a practical scenario where noise is introduced into the original data. As a result, the data we are analyzing is presented as follows
\begin{equation}
    \label{tilde xy}
    \left\{\begin{aligned}
        \tilde{x}[n] &= x_d[n] + \sigma_1 \zeta_1[n] \\
        \tilde{y}[n] &= y_d[n] + \sigma_2 \zeta_2[n]
    \end{aligned}\right.
\end{equation}
where $\zeta_1$ and $\zeta_2$ are two independent stochastic processions obeying the standard normal distribution, $\sigma_1$ and $\sigma_2$ are two positive constants denoting the standard errors of the Gauss noise. \\
\indent Thus, the path data could be viewed as a combination of the accurate data and a Gauss noise. So this series is a stochastic procession. Owning to the reason there is no available information about the original path, so it is nearly impossible to construct a perfect guiding vector field like \ref{vector field expression}. Thus, an approximation $\hat{x}(\theta)$ and $\hat{y}(\theta)$ is necessary, which would be given later. Here we assume the approximation has been given. Then, the reconstructed curve is
\begin{equation}
    \label{hat phi}
    \left\{\begin{aligned}
        \hat{\phi}_1 &= x(t) - \hat{x}(\theta)\\
        \hat{\phi}_2 &= y(t) - \hat{y}(\theta)
    \end{aligned}\right.
\end{equation}
and the corresponding guiding vector field is
\begin{equation}
    \label{hat vector field expression}
    \hat{\chi} = \begin{bmatrix}
        -\frac{\partial \hat{\phi}_1}{\theta} \\ -\frac{\partial \hat{\phi}_2}{\theta} \\ 1
    \end{bmatrix}
    -k_1 \hat{\phi}_1 \begin{bmatrix}
        1 \\ 0 \\ \frac{\partial \hat{\phi}_1}{\theta}
    \end{bmatrix}
    -k_2 \hat{\phi}_2 \begin{bmatrix}
        0 \\ 1 \\ \frac{\partial \hat{\phi}_2}{\theta}
    \end{bmatrix}
\end{equation}
For a stochastic procession, the mean square error is often considered:
\begin{align}
    \label{mean square error}
        P &= E\left\{[x_d(\theta) - \hat{x}(\theta)]^2 + [y_d(\theta) - \hat{y}(\theta)]^2\right\} \nonumber\\
        &= \int_0^{2\pi}\left \{[x_d(\theta) - \hat{x}(\theta)]^2 + [y_d(\theta) - \hat{y}(\theta)]^2\right\} \mathrm{d} \theta
\end{align}
For convenience, define $[x_d(\theta) - \hat{x}(\theta)]^2 + [y_d(\theta) - \hat{y}(\theta)]^2$ as $p(\theta)$. \\
\indent While in a path following problem, we usually concern more about the path-following error, which can be written as
\begin{align}
    \label{pf error}
        e(\theta(t),t) =&~ [x(t) - x_d(\theta)]^2 + [y(t) - y_d(\theta)]^2 \nonumber\\
        =& ~[x(t) -\hat{x}(\theta)+ \hat{x}(\theta) - x_d(\theta)]^2  \nonumber\\
        & ~+ [y(t) -\hat{y}(\theta)+\hat{y}(\theta)- y_d(\theta)]^2 \nonumber\\
        \leq &~[x(t) -\hat{x}(\theta)]^2+ [\hat{x}(\theta) - x_d(\theta)]^2 \nonumber\\
        & ~+ [y(t) -\hat{y}(\theta)]^2+ [\hat{y}(\theta)- y_d(\theta)]^2 \nonumber\\
        =&~ \hat{\phi}_1^2(\theta,t)+\hat{\phi}_2^2(\theta,t) + p(\theta)
\end{align}
where $\hat{\phi}_1:=x(t) -\hat{x}(\theta)$ and $\hat{\phi}_2:=y(t) -\hat{y}(\theta)$. It is noteworthy that \emph{Theorem} \ref{theorem1} has shown that $\hat{\phi}_1^2+\hat{\phi}_2^2$ would converge to zero asymptotically. So the final accumulation of error is exactly (\ref{mean square error}). To restrict the impact of Gauss noise, we need to introduce a window function as a filter. However, the requirement of this mechanism is self-conflicted:
\begin{itemize}
    \item In order to enhance the accuracy of reconstructing the initial path, it is imperative that the major information need to be retained.
    \item To attenuate the noise, it is necessary to exclude some information.
\end{itemize}
\indent To balance this contradiction, a rectangular window is introduced herein. Based on the Fourier series $\tilde{a}_k$ from $\tilde{c}[n] = \tilde{x}[n] + \mathrm{j}\tilde{y}[n]$, the approximation coefficients are
\begin{equation}
    \label{window function}
    \hat{a}_k = w[k]\tilde{a}_k,\ w[k] = \left\{ \begin{aligned}
        &1, k\in \bar{Z}_m \\
        &0, \text{otherwise}
    \end{aligned} \right.
\end{equation}
This window function works as Fig. \ref{fig3} shows. It should be emphasized that the Fourier spectrum of a normally distributed variable is a constant.

Hence, for larger $m$, $\hat{a}_k$ would remain more information. On the contrary, $\hat{a}_k$ would remain less information for smaller $m$. The approximation of the original curve is
\begin{equation}
    \label{hatxy}
    \left\{\begin{aligned}
        \hat{x}(\theta) &= \sum_{k \in \bar{Z}_m} \Vert \hat{a}_k \Vert \cos{(k\theta + \mathrm{arg}(\hat{a}_k))} \\
        \hat{y}(\theta) &= \sum_{k \in \bar{Z}_m} \Vert \hat{a}_k \Vert \sin{(k\theta + \mathrm{arg}(\hat{a}_k))}
    \end{aligned}\right.
\end{equation}
Then,
\begin{equation*}
    \label{p}
    \begin{aligned}
        p(\theta) &= [x_d(\theta) - \hat{x}(\theta)]^2 + [y_d(\theta) - \hat{y}(\theta)]^2 \\
        &= \Vert c_d(\theta) - \hat{c}(\theta)\Vert^2 \\
        &= \left\| \sum_{k \in \bar{Z}_N} a_k e^{\mathrm{j}k\theta} -  \sum_{k \in \bar{Z}_m} \hat{a}_k e^{\mathrm{j}k\theta} \right\|^2 \\
        &= \left\| \sum_{k \in \bar{Z}_m} (a_k-\hat{a}_k) e^{\mathrm{j}k\theta} + \sum_{k \in \bar{Z}_N \setminus \bar{Z}_m} a_k e^{\mathrm{j}k\theta} \right\|^2 \\
        &\leq \left\| \sum_{k \in \bar{Z}_m} (a_k-\hat{a}_k) e^{\mathrm{j}k\theta} \right\|^2 + \left\| \sum_{k \in \bar{Z}_N \setminus \bar{Z}_m} a_k e^{\mathrm{j}k\theta}\right\|^2 \\
    \end{aligned}
\end{equation*}
Define these two terms as
\begin{align*}
    p_1 &= \left\Vert \sum_{k \in \bar{Z}_m} (a_k-\hat{a}_k) e^{\mathrm{j}k\theta}\right\Vert^2, \\
    p_2 &= \left\Vert \sum_{k \in \bar{Z}_N \setminus \bar{Z}_m} a_k e^{\mathrm{j}k\theta} \right\Vert^2.
\end{align*}
and then we have
\begin{equation}
    \label{p2}
    p_2 \leq  \sum_{k \in \bar{Z}_N \setminus \bar{Z}_m} \Vert a_k  \Vert^2
\end{equation}
and
\begin{equation}
    \label{p1}
    \begin{aligned}
        p_1 &= \frac{1}{N^2}\left\| \sum_{k \in \bar{Z}_m} (\sum_{l \in \bar{Z}_N} (c[l]-\tilde{c}[l])e^{\mathrm{j}l\theta})e^{\mathrm{j}k\theta} \right\|^2 \\
        &= \frac{1}{N^2} \left\| \sum_{k \in \bar{Z}_m} (\sum_{l \in \bar{Z}_N} (\zeta_1[l]+j \zeta_2[l])e^{\mathrm{j}l\theta})e^{\mathrm{j}k\theta} \right\|^2
    \end{aligned}
\end{equation}
Noticing the fact that $\zeta_1$ and $\zeta_2$ are two independent variable following the normal distribution. From a mean-square perspective,
\begin{equation}
    \label{ms p1}
        E(p_1) \leq 2\pi \frac{m^2}{N^2}(\sigma_1^2+\sigma_2^2)
\end{equation}
Thus,
\begin{align}
    \label{Pm}
        P(m) &= E(p) \nonumber\\
        &\leq 2\pi \frac{m^2}{N^2}(\sigma_1^2+\sigma_2^2) + 2\pi \sum_{k \in \bar{Z}_N \setminus \bar{Z}_m} \Vert a_k  \Vert^2 \nonumber\\
        &:=\bar{P}(m)
\end{align}
which provides an upper bound for the mean-square error. If we define $F(m) = \bar{P}(m)-\bar{P}(m-1)$ for $m\geq 2$ as a backward differential for the upper bound of the mean-square error, then
\begin{equation}
    \label{Fm}
    F(m) = 2\pi\Big[\frac{2m-1}{N^2} (\sigma_1^2+\sigma_2^2)-\Vert a_{m^*} \Vert^2\Big]
\end{equation}
where $m^* \in \bar{Z}_N\setminus \bar{Z}_m$ while  $m^* \notin \bar{Z}_N\setminus \bar{Z}_{m-1}$. Equation (\ref{Fm}) indicates that the error part contributed by the Gauss noise increases as $m$ get larger while the error part contributed by the original data decreases.

\begin{Remark}
    Inequality (\ref{Pm}) is valid due to the fact that the Fourier spectrum of a normally distributed variable is constant, whereas for a series of specific data, the spectrum is typically not constant. The use of rectangular windows can function as a filter to effectively counteract the presence of noise. Additionally, the second term in (\ref{Pm}) indicates that in the case of curves with complex boundaries, such as fractal boundaries, it is exceedingly difficult to accurately reconstruct the boundary while simultaneously mitigating the substantial impact of Gaussian noise as there is too much information contained in high-frequency domain.
\end{Remark}

\indent An additional noteworthy observation is that, when Gaussian noise is added to a specific set of data, it is possible to identify a particular value of $m$ that minimizes the upper bound for $P(m)$. However, in practical applications, the selection of $m$ must not only aim to reduce errors but also account for the computational capacity of the hardware. Hence, the selection of variable $m$ should aim to minimize the upper bound of the mean square error as much as possible within the computational capacity.

This section would be ended with the answer to the Problem $1$. Here is the mean square error for path following could be defined as
\begin{equation}
    \label{mspferror}
    e_{ms} (t) = \int_{0}^{2\pi}e(\theta(t),t)d\theta
\end{equation}
(\ref{pf error}) and (\ref{Pm}) imply the mean square path following error has an ultimate upper bound given by (\ref{Pm}). This ultimate upper bound could be donated as $\delta(m)$. Thus, the second requirement of Problem 1 holds.

\section{Simulation}
\label{section4}
\begin{figure*} [t!]
	\centering
	\subfloat[\label{fig:4a}]{
		\includegraphics[width=0.25 \linewidth]{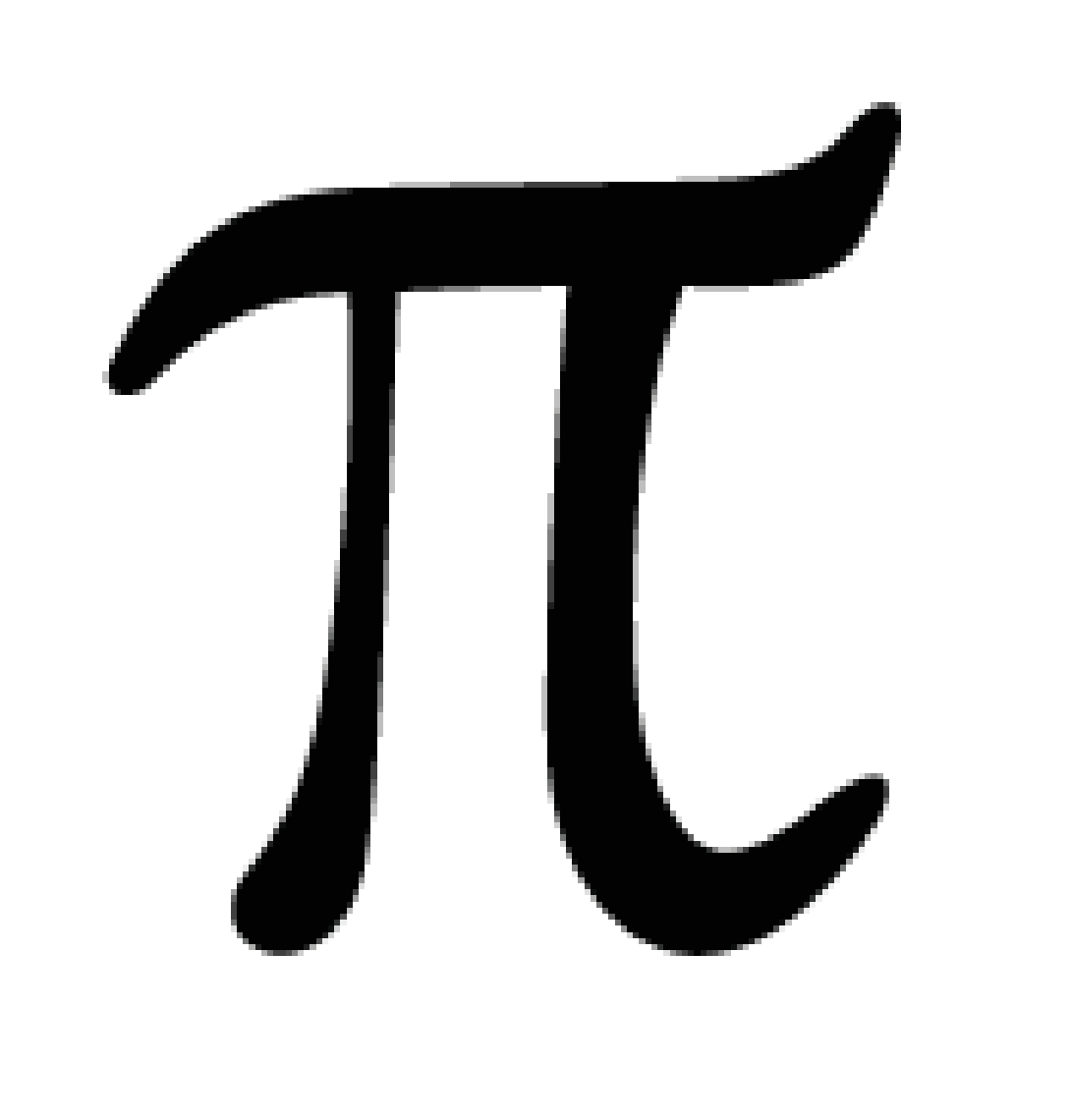}}
	\subfloat[\label{fig:4b}]{
		\includegraphics[width=0.33 \linewidth]{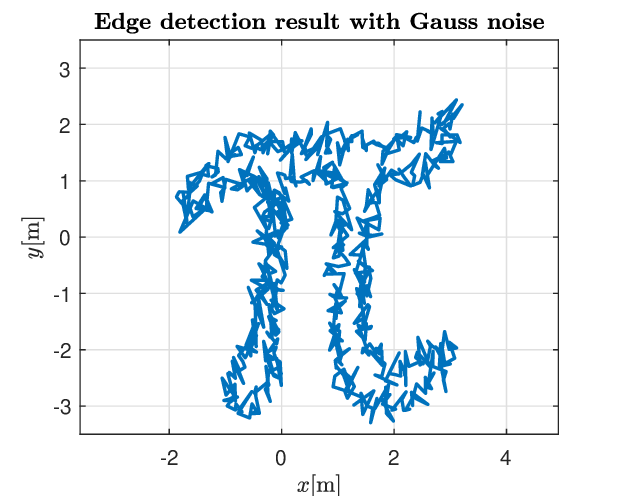}}
    \subfloat[\label{fig:4c}]{
		\includegraphics[width=0.33 \linewidth]{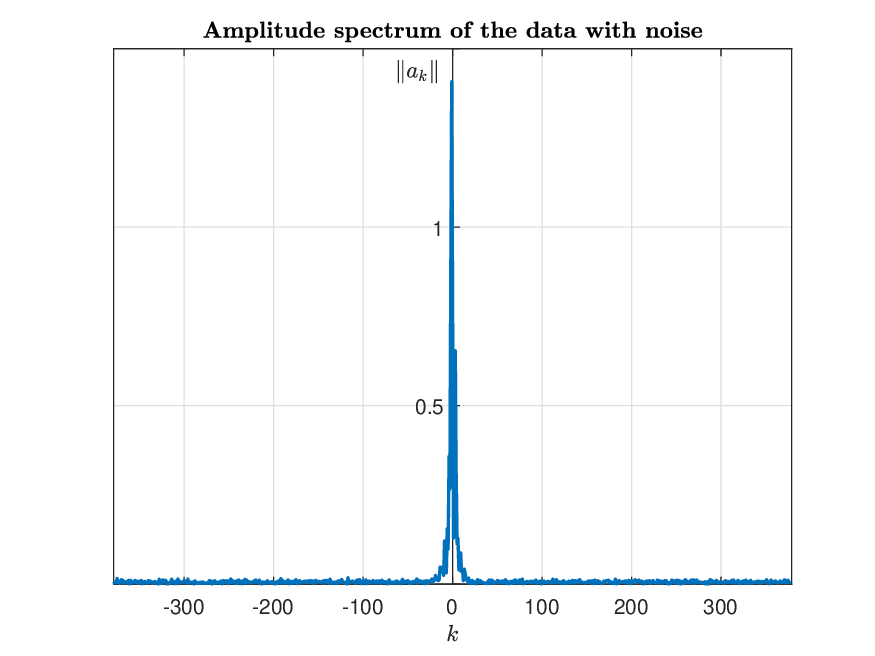}}
	\\
	\subfloat[\label{fig:4d}]{
		\includegraphics[width=0.33 \linewidth]{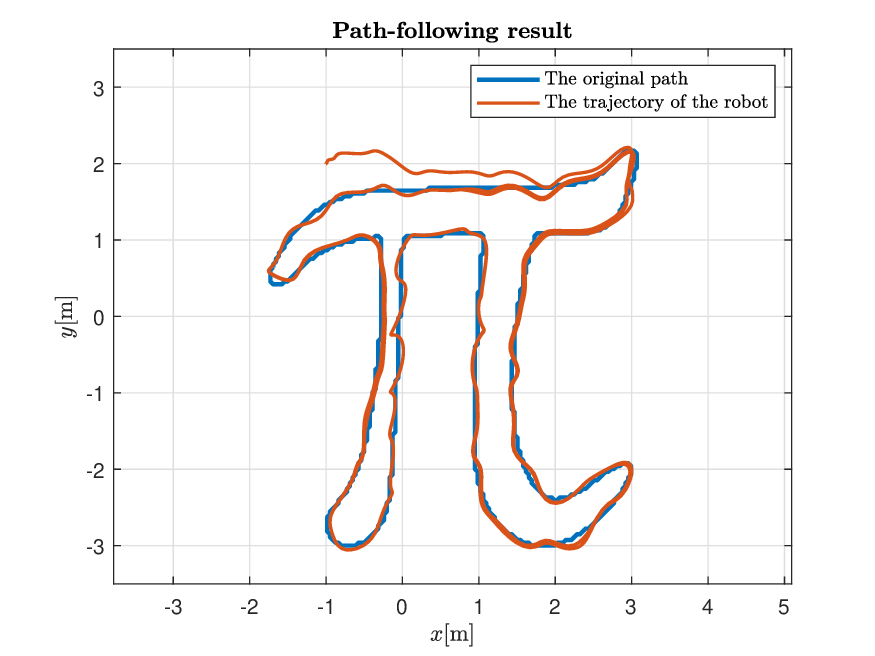}}
	\subfloat[\label{fig:4e}]{
		\includegraphics[width=0.33 \linewidth]{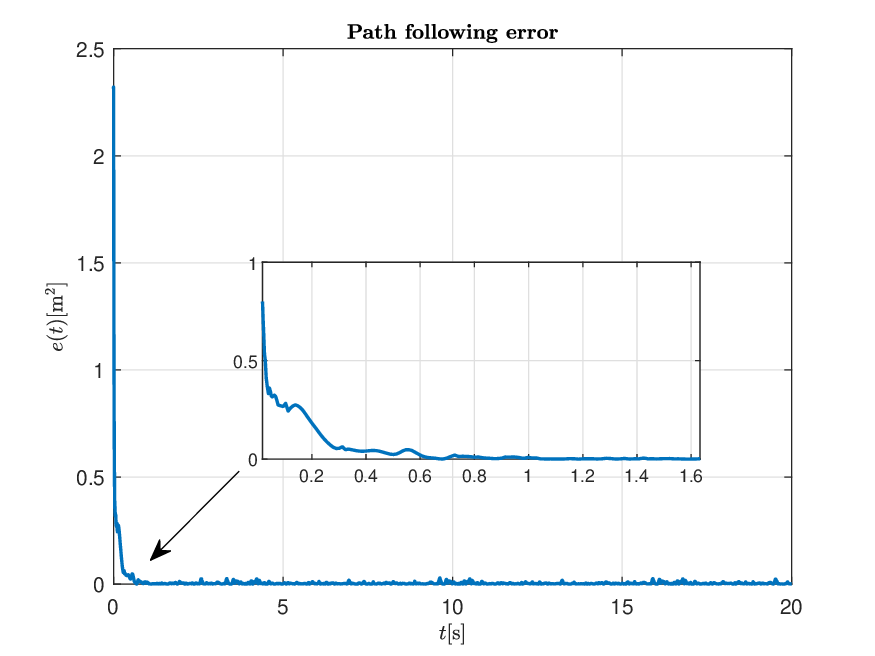}}
    \subfloat[\label{fig:4f}]{
        \includegraphics[width=0.33 \linewidth]{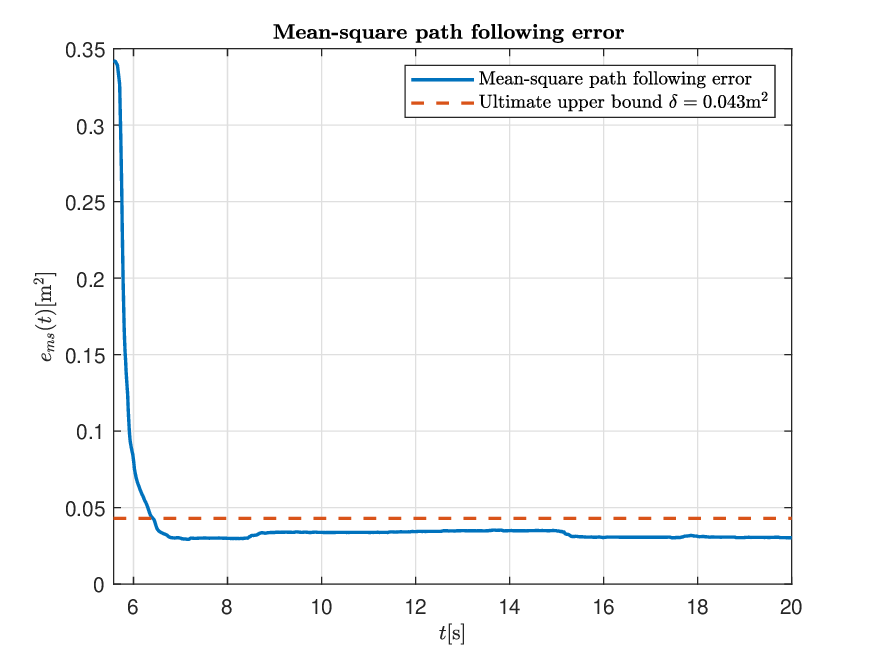}}
	\caption{Simulation Results. (\ref{fig:4a}) the original pattern. (\ref{fig:4b}) Edge detection result with Gauss noise. (\ref{fig:4c}) Edge Amplitude spectrum of the data with noise. (\ref{fig:4d}) Path-following result. (\ref{fig:4e}) Path following error. (\ref{fig:4f}) Mean-square path following error. }
	\label{fig4}
\end{figure*}

To demonstrate the effectiveness of the proposed method, a numerical simulation is presented in this section. The desired path follows a $\pi$-like pattern, and is obtained from an edge detection algorithm applied to the original image, resulting in a set of $758$ discrete data points. Two independent noise signals, each with a normally distributed standard deviation of $\sigma_1 = 0.1\mathrm{m}$ and $\sigma_2 = 0.15\mathrm{m}$, respectively, are added to the data. The rectangular window used for processing the data had a width of $m = 100$. The vector field parameters are chosen as $k_1 = 1$ and $k_2 = 1$. The robot trajectory is governed by the dynamic system $\dot{\xi}(t) = \chi(\xi(t))$, with an initial condition $[-1,2]^\top$. The simulation duration is $20s$. The upper bound for the ultimate error in this simulation is determined to be $\delta = 0.043\mathrm{m}^2$. The simulation results are presented in Fig. \ref{fig4}, which shows that the trajectory of the robot could track the original path with a mean square error less than $\delta$.
\section{Conclusion}
\label{section5}
The paper presents a path tracking algorithm that can work with discrete data and is globally relevant. It has the ability to reduce measurement errors and can estimate the maximum mean squared tracking error. The algorithm uses fast Fourier transforms for data processing, and the width of the window function can be controlled to adjust the computational complexity and tracking precision.

This article suggests some possible directions for future research. In order to reduce computational complexity, a rectangular window function was directly employed in this study. However, this window function may not be ideal for suppressing specific types of noise. If computational resources allow, it is feasible to use other window functions to further reduce the mean square error. In addition, based on edge detection algorithms, obstacle edges can be easily obtained. Combining this with the method proposed in this article, a practical obstacle avoidance algorithm may be developed.

\IEEEpeerreviewmaketitle

\bibliographystyle{IEEEtran}
\bibliography{references}

\begin{thebibliography}{10}
\providecommand{\url}[1]{#1}
\csname url@samestyle\endcsname
\providecommand{\newblock}{\relax}
\providecommand{\bibinfo}[2]{#2}
\providecommand{\BIBentrySTDinterwordspacing}{\spaceskip=0pt\relax}
\providecommand{\BIBentryALTinterwordstretchfactor}{4}
\providecommand{\BIBentryALTinterwordspacing}{\spaceskip=\fontdimen2\font plus
\BIBentryALTinterwordstretchfactor\fontdimen3\font minus
  \fontdimen4\font\relax}
\providecommand{\BIBforeignlanguage}[2]{{%
\expandafter\ifx\csname l@#1\endcsname\relax
\typeout{** WARNING: IEEEtran.bst: No hyphenation pattern has been}%
\typeout{** loaded for the language `#1'. Using the pattern for}%
\typeout{** the default language instead.}%
\else
\language=\csname l@#1\endcsname
\fi
#2}}
\providecommand{\BIBdecl}{\relax}
\BIBdecl

\bibitem{Yao_2020Path}
W.~Yao and M.~Cao, ``\BIBforeignlanguage{en}{Path following control in 3d using
  a vector field},'' \emph{\BIBforeignlanguage{en}{Automatica}}, vol. 117, p.
  108957, 2020.

\bibitem{Wang_2019Cooperative}
Y.~Wang, D.~Wang, and S.~Zhu, ``\BIBforeignlanguage{en}{Cooperative moving path
  following for multiple fixed-wing unmanned aerial vehicles with speed
  constraints},'' \emph{\BIBforeignlanguage{en}{Automatica}}, vol. 100, pp.
  82--89, 2019.

\bibitem{Oliveira_2016}
T.~Oliveira, A.~P. Aguiar, and P.~Encarnacao, ``\BIBforeignlanguage{en}{Moving
  path following for unmanned aerial vehicles with applications to single and
  multiple target tracking problems},'' \emph{\BIBforeignlanguage{en}{IEEE
  Transactions on Robotics}}, vol.~32, no.~5, pp. 1062--1078, 2016.

\bibitem{Zuo_2022Coordinated}
Z.~Zuo, J.~Song, and Q.-L. Han, ``\BIBforeignlanguage{en}{Coordinated planar
  path-following control for multiple nonholonomic wheeled mobile robots},''
  \emph{\BIBforeignlanguage{en}{IEEE Transactions on Cybernetics}}, vol.~52,
  no.~9, pp. 9404--9413, 2022.

\bibitem{Sujit_2014Unmanned}
P.~Sujit, S.~Saripalli, and J.~B. Sousa, ``\BIBforeignlanguage{en}{Unmanned
  aerial vehicle path following: A survey and analysis of algorithms for
  fixed-wing unmanned aerial vehicless},'' \emph{\BIBforeignlanguage{en}{IEEE
  Control Systems}}, vol.~34, no.~1, pp. 42--59, 2014.

\bibitem{Goncalves_2010}
V.~M. Goncalves, L.~C.~A. Pimenta, C.~A. Maia, B.~C.~O. Dutra, and G.~A.~S.
  Pereira, ``\BIBforeignlanguage{en}{Vector fields for robot navigation along
  time-varying curves in $n$-dimensions},'' \emph{\BIBforeignlanguage{en}{IEEE
  Transactions on Robotics}}, vol.~26, no.~4, pp. 647--659, 2010.

\bibitem{Yao_2021Singularity}
W.~Yao, H.~G. de~Marina, B.~Lin, and M.~Cao,
  ``\BIBforeignlanguage{en}{Singularity-free guiding vector field for robot
  navigation},'' \emph{\BIBforeignlanguage{en}{IEEE Transactions on Robotics}},
  pp. 1--16, 2021.

\bibitem{Rezende_2022Constructive}
A.~M.~C. Rezende, V.~M. Goncalves, and L.~C.~A. Pimenta,
  ``\BIBforeignlanguage{en}{Constructive time-varying vector fields for robot
  navigation},'' \emph{\BIBforeignlanguage{en}{IEEE Transactions on Robotics}},
  vol.~38, no.~2, pp. 852--867, 2022.

\bibitem{Yao_2023Topological}
W.~Yao, B.~Lin, B.~D.~O. Anderson, and M.~Cao,
  ``\BIBforeignlanguage{en}{Topological analysis of vector-field guided path
  following on manifolds},'' \emph{\BIBforeignlanguage{en}{IEEE Transactions on
  Automatic Control}}, vol.~68, no.~3, pp. 1353--1368, 2023.

\bibitem{9373984}
Q.~Zhang and H.~Wang, ``A novel data-based stochastic distribution control for
  non-gaussian stochastic systems,'' \emph{IEEE Transactions on Automatic
  Control}, vol.~67, no.~3, pp. 1506--1513, March 2022.

\bibitem{9780596}
X.~Zhao and F.~Deng, ``Sufficient and necessary condition for the asymptotic
  stability of stochastic systems with discrete time feedbacks and
  applications,'' \emph{IEEE Transactions on Automatic Control}, vol.~67,
  no.~9, pp. 4717--4732, Sep. 2022.

\bibitem{9362158}
------, ``Time-varying halanay inequalities with application to stability and
  control of delayed stochastic systems,'' \emph{IEEE Transactions on Automatic
  Control}, vol.~67, no.~3, pp. 1226--1240, March 2022.

\bibitem{9484756}
J.~G. Rueda-Escobedo, E.~Fridman, and J.~Schiffer, ``Data-driven control for
  linear discrete-time delay systems,'' \emph{IEEE Transactions on Automatic
  Control}, vol.~67, no.~7, pp. 3321--3336, July 2022.

\bibitem{10101826}
H.~Modares, ``Data-driven safe control of uncertain linear systems under
  aleatory uncertainty,'' \emph{IEEE Transactions on Automatic Control}, pp.
  1--8, 2023.

\bibitem{10122597}
H.~van Waarde, J.~Eising, M.~Camlibel, and H.~Trentelman, ``A behavioral
  approach to data-driven control with noisy input-output data,'' \emph{IEEE
  Transactions on Automatic Control}, pp. 1--14, 2023.

\bibitem{10107903}
Y.~Li, S.~X. Ding, G.~Liu, and C.~Hua, ``Sampled-data based distributed output
  feedback leader-following consensus for time-delay multiagent systems,''
  \emph{IEEE Transactions on Automatic Control}, pp. 1--8, 2023.

\bibitem{9345499}
M.~Sharf, A.~Koch, D.~Zelazo, and F.~Allgöwer, ``Model-free practical
  cooperative control for diffusively coupled systems,'' \emph{IEEE
  Transactions on Automatic Control}, vol.~67, no.~2, pp. 754--766, Feb 2022.

\bibitem{khalil2002nonlinear}
H.~Khalil, \emph{Nonlinear Systems}, ser. Pearson Education.\hskip 1em plus
  0.5em minus 0.4em\relax Prentice Hall, 2002.

\end{thebibliography}

\begin{IEEEbiography}[{\includegraphics[width=1in,height=1.25in,clip,keepaspectratio]{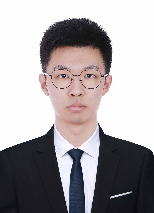}}]{Zirui Chen} received the B.Eng. degree in automation from Beihang University (BUAA), Beijing, China, in 2021, where he is currently pursuing the M.A. Eng. degree in control theory and applications. 

    His research interests are in the fields of nonlinear system control and geometric control.
\end{IEEEbiography}

\begin{IEEEbiography}[{\includegraphics[width=1in,height=1.25in,clip,keepaspectratio]{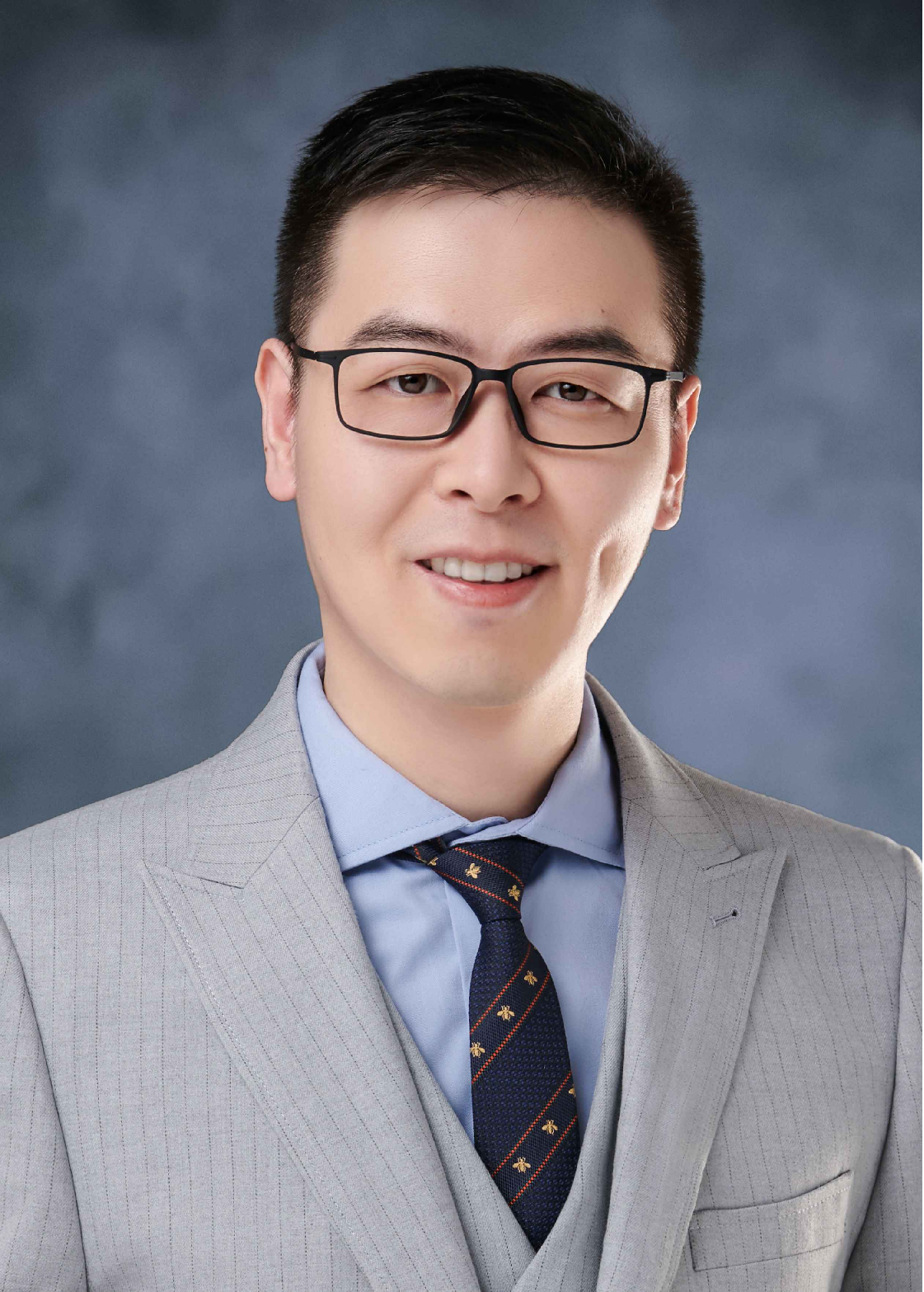}}]{Zongyu Zuo} (Senior Member, IEEE) received his B.Eng. degree in Automatic Control from Central South University, Hunan, China, in 2005, and Ph.D. degree in Control Theory and Applications from Beihang University (BUAA), Beijing, China, in 2011.

He was an academic visitor at the School of Electrical and Electronic Engineering, University of Manchester from September 2014 to September 2015 and held an inviting associate professorship at Mechanical Engineering and Computer Science, UMR CNRS 8201, Universit\'{e} de Valenciennes et du Hainaut-Cambr\'{e}sis in October 2015 and May 2017. He is currently a full professor at the School of Automation Science and Electrical Engineering, Beihang University. His research interests are in the fields of nonlinear system control, control of UAVs, and coordination of multi-agent system. He was identified as a Highly Cited Researcher - 2020 and 2022 by Clarivate Analytics as well as a most cited Chinese Researcher - 2021 and 2022  by Elsevier.

Prof. Zuo currently serves as an Associate Editor for IEEE Transactions on Industrial Informatics, IEEE/CAA Journal of Automatica Sinica, Journal of The Franklin Institute, Journal of Vibration and Control, and International Journal of Aeronautical {\&} Space Sciences.
\end{IEEEbiography}

\end{document}